\documentclass[aps,prb,twocolumn,letterpaper]{revtex4}

\usepackage{graphicx}
\usepackage{dcolumn}
\usepackage{bm}
\usepackage{longtable}
\usepackage{amsfonts,amsmath,amssymb}

\newcommand{\BFCO}{Bi$_2$FeCrO$_6$}
\newcommand{\BFO}{BiFeO$_3$}
\newcommand{\BCO}{BiCrO$_3$}

\begin{document}

\title{First principles study of the multiferroics \BFO, \BFCO, and
  \BCO: Structure, polarization, and magnetic ordering temperature}

\author{Pio Baettig} 
\email{baettig@mrl.ucsb.edu} 
\altaffiliation[Also at ]{Chemistry Department, Universit\'{e} de
Fribourg, P\'{e}rolles, CH-1700 Fribourg, Switzerland.}
\author{Claude Ederer} 
\author{Nicola A. Spaldin}
\affiliation{Materials Department and Materials Research Laboratory,
University of California, Santa Barbara, California 93106, USA}

\date{\today}

\begin{abstract}
We present results of an {\it ab initio} density functional theory
study of three bismuth-based multiferroics, \BFO, \BFCO\ and \BCO. We
disuss differences in the crystal and electronic structure of the
three systems, and we show that the application of the LDA+$U$ method
is essential to obtain realistic structural parameters for \BFCO. We
calculate the magnetic nearest neighbor coupling constants for all
three systems and show how Anderson's theory of superexchange can be
applied to explain the signs and relative magnitudes of these coupling
constants. From the coupling constants we then obtain a mean-field
approximation for the magnetic ordering temperatures. Guided by our
comparison of these three systems, we discuss the possibilities for
designing a multiferroic material with large magnetization above room
temperature.
\end{abstract}

\pacs{Valid PACS appear here}
\maketitle

\section{\label{sec:intro}Introduction}

There has been growing recent interest in the magnetic, ferroelectric,
and piezoelectric properties of bismuth-based perovskite-structure
oxides.\cite{Hill_Rabe,Seshadri_Hill,Hill_Pio,Wang/Neaton_Science:2003,BiCrO3_Niitaka,Baettig2005a,Neaton_et_al:2005,Ederer/Spaldin:2005,Bai_et_al:2005,Baettig2005}
In such materials, the formally Bi$^{3+}$ ion occupies the perovskite
$A$ site, and its stereochemically active $6s^2$ lone pair induces a
symmetry-lowering structural distortion that can lead to
ferroelectricity.\cite{Seshadri_Hill} Part of this interest lies in
increasing concern about the environmental toxicity of lead-based
piezoelectrics and the resulting search for alternative
materials.\cite{Baettig2005,JRCheng1}
In addition, the stereochemical activity of Bi is being exploited to
induce structural distortions in magnetic oxides, with the goal of
forming ferromagnetic ferroelectrics (so-called magnetoelectric
multiferroics).\cite{Hill_Rabe,Seshadri_Hill,Hill_Pio,Wang/Neaton_Science:2003,BiCrO3_Niitaka,Baettig2005a,Neaton_et_al:2005,Ederer/Spaldin:2005,Bai_et_al:2005}
Although a large number of potential applications can be envisaged for
a material which is simultaneously ferromagnetic and
ferroelectric,\cite{Wood_Austin} there are currently no known
single-phase materials which show large, robust magnetization and
polarization at room temperature.

In this work we analyze the magnetic properties of three related
Bi-based multiferroics in order to understand the chemical and
structural factors which affect the magnetic ordering temperature. Our
goal is to develop guidelines for designing new perovskite-structure
ferroelectrics with a large macroscopic magnetization above room
temperature.  The focus of our investigation is the (111) layered
double perovskite {\BFCO}, and its parent compounds \BFO\ and
\BCO. \BFO\ is the most well-studied of the three compounds and has
long been known to be ferroelectric, with a Curie temperature $T_C
\sim 1100$~K, and antiferromagnetic, with a N\'{e}el temperature $T_N
\sim 640$~K.\cite{Kiselev_BFO_AFM:1963,Teague} It crystallizes in a
rhombohedrally distorted perovskite structure with space group
$R3c$.\cite{Michel_BFO_structure:1969} Recently, interest in this
material has increased considerably, due to the report of a large
electric polarization and a non-zero magnetization in epitaxial films
of \BFO, both above room temperature.\cite{Wang/Neaton_Science:2003}
The large electric polarization, as well as the small magnetization
found in 200-400~nm thick \BFO\ films, have been confirmed by other
experiments and by first principles calculations
\cite{Bai_et_al:2005,Neaton_et_al:2005,Ederer/Spaldin:2005} (see also
Section~\ref{pol}), whereas the origin of the large magnetization
observed for film thickness $<$ 100~nm is still under debate.

\BCO\ was recently synthesized and reported to be a highly distorted
perovskite with $C2$ symmetry;\cite{BiCrO3_Niitaka} a polar space
group which permits the occurrence of ferroelectricity.
Antiferromagnetic ordering was reported with a residual magnetization
of 0.02 $\mu_{B}$ per Cr, suggestive of weak ferromagnetism.


\BFCO~ has not yet been realized experimentally, but has been
predicted theoretically to be ferrimagnetic (with a magnetic moment of
2 $\mu_{B}$ per formula unit) and ferroelectric (with a polarization
of $\sim$ 80~$\mu$C/cm$^{2}$).\cite{Baettig2005a} The predicted ground
state structure is very similar to the $R3c$ structure of \BFO, except
that in every second (111) layer the Fe$^{3+}$ cations are replaced by
Cr$^{3+}$, which reduces the symmetry to the space group $R3$. Such a
$B$-site ordered structure could be realized by (111) layer-by-layer
growth on an appropriate substrate.\cite{Ueda1998}

The remainder of this article is structured as follows: First we
present the computational details of our calculations. Then, in
Section \ref{sec:res} A-C, we compare the calculated ground state
structures of the three systems, with a particular focus on the
influence of the exchange-correlation potential used in the density
functional formalism on the structural parameters; this is
particularly important for reproducing the correct physics in these
strongly-correlated magnetic insulators. We also compare the
calculated electric polarization of the three materials. In Section
\ref{section:M_T} we calculate nearest and next-nearest neighbor
magnetic couplings and determine the magnetic ordering temperatures
for the three materials using the mean-field approximation. Finally,
in Section~\ref{discussion} we discuss the variation of the nearest
neighbor couplings in these systems and the implications of our
results for the design of new multiferroics with higher magnetic Curie
temperatures.

\section{Computational Details}
\label{methods}

All calculations described in this work were performed using the
projector augmented wave (PAW) formalism \cite{bloechlPAW} of density
functional theory,\cite{hoh64,koh65} implemented in the \emph{Vienna
Ab-Initio Simulation Package}
(VASP).\cite{Kresse/Furthmueller_VASP2:1996,KressePAW} For BiFeO$_3$
the Bi 5$d$ and the Fe 3$p$ semicore states were treated as valence
electrons, for Bi$_2$FeCrO$_3$ and BiCrO$_3$ we did not include
semicore states in the valence. The local spin density approximation
(LDA) as well as the LDA+$U$ method in the so-called fully localized
limit\cite{Anisimov:1997} was used to treat exchange and
correlation. Except where otherwise noted, the structural parameters
were fixed to those obtained by full structural optimization using
$U$=3~eV and $J$=0.8~eV (for \BFCO\ and \BCO) or $J$=1~eV (for \BFO)
for the treatment of the transition metal $d$ orbitals. Then, the
electronic structure and magnetic coupling constants were calculated
as a function of $U$ (and fixed $J$) in the range $U$=3~eV - 6~eV. For
simplicity, the same $U$ and $J$ were used on both the Fe and Cr sites
in \BFCO. All our results are well converged with respect to k-point
mesh and the energy cutoff for the plane wave expansion.  For details
regarding the structural optimization of \BFO\ see
Ref.~\onlinecite{Neaton_et_al:2005}, and
Ref.~\onlinecite{Baettig2005a} for \BFCO. The electric polarization
was calculated using the Berry phase method.\cite{kv,vanderbilt93}

\section{\label{sec:res} Results and discussion}

\subsection{\label{subsec:GS}Ground state structures and magnetic orderings}

We begin by comparing the calculated structures of the three
compounds. The ground state structure of \BFCO~ was calculated in
Ref.~\onlinecite{Baettig2005a} by optimizing the geometries of a range
of starting configurations, obtained by freezing in high-symmetry
unstable phonon modes of BiCrO$_3$ \cite{Hill_Pio} and
BiAlO$_3$.\cite{Baettig2005} The energies of structures constrained to
the resulting symmetries are listed in Table~\ref{table:FMhs40} for
two magnetic configurations: ferromagnetic, and the ferrimagnetic
equivalent of so-called G-type antiferromagnetic ordering, in which
all spins within the same (111) plane are ferromagnetically aligned,
with antiparallel alignment of spins in adjacent (111) planes. We see
that the lowest energy structure of all the combinations tried has
$R3$ symmetry (space-group 146), with alternating rotations of the
oxygen octahedra around the [111] direction, combined with relative
displacements of all ions along [111]. This structure is closely
related to the $R3c$ structure found experimentally for
BiFeO$_3$,\cite{Michel_BFO_structure:1969} but with an additional
symmetry lowering due to the different $B$ cations. From
Table~\ref{table:FMhs40} one can also see that the ferrimagnetic
configuration has a lower energy than the ferromagnetic case for all
structural symmetries. The net magnetic moment of the ferrimagnetic
case is 2 $\mu_B$ per Fe-Cr pair, which corresponds to a magnetization
of 160.53 emu/cm$^3$ in the case of the ground state $R3$ structure.
It is also clear from Table~\ref{table:FMhs40} that typical energy
differences between different magnetic configurations are
significantly smaller than the energy differences due to different
structural symmetries. Moreover, the structural parameters (not shown)
obtained for FM and FiM ordering within a given structural symmetry
differ only slightly.

\begin{table}
\begin{center}
\caption{Total energies per formula unit of Bi$_2$FeCrO$_6$, for different
structural symmetries relative to the ground state structure for ferromagnetic
(FM) and ferrimagnetic (FiM) ordering, respectively.
\label{table:FMhs40}}
\begin{ruledtabular}
\begin{tabular}{llc}
space group & magnetic ordering & $E$ [eV/f.u.] \\ 
\hline \\[-3mm]
$Pm\bar{3}m$ & FM  & 1.722 \\ 
             & FiM & 1.670 \\ 
\hline 
$Cmca$       & FM  & 1.215 \\ 
             & FiM & 1.096 \\ 
\hline 
$P4/mnc$     & FM  & 0.837 \\ 
             & FiM & 0.683 \\ 
\hline
$R3m$        & FM  & 0.612 \\ 
             & FiM & 0.473 \\ 
\hline 
$R3$         & FM  & 0.176 \\ 
             & FiM & 0.000 \\
\end{tabular}
\end{ruledtabular}
\end{center}
\end{table}

To exclude the existence of a more complicated magnetic configuration with
lower energy, we doubled the size of the unit cell along one of the
rhombohedral lattice vectors and compared the total energies for all possible
collinear spin configurations within the resulting supercell. The doubled unit
cell allows for spin configurations with antiparallel alignment of magnetic
cations of the same type, which would lead to a cancellation of the
macroscopic magnetization. For these calculations no additional structural
relaxations were performed; the structure was fixed to that obtained by
relaxation within the ferrimagnetic $R3$ symmetry. Based on the very similar
structures we obtained above for FM and FiM orderings, we do not expect that
further structural relaxation would alter the relative energies of the
different spin configurations.  It was found that the ``G-type-like''
ferrimagnetic configuration described above is indeed the ground state, and no
cancellation of the macroscopic magnetization occurs. We did not investigate
the effect of spin-orbit coupling and the possibility of noncollinear spin
structures or long-wavelength magnetic ordering.

For \BFO\ both the structure and the magnetic configuration are well
established.\cite{Michel_BFO_structure:1969,Kubel/Schmid_BFO_structure:1990}
Therefore, we did not perform a similar thorough ground state search
for this system and instead only relaxed the structure within the
experimentally found $R3c$ symmetry in combination with G-type
antiferromagnetic ordering. The details of the structural relaxation
of \BFO\ have already been presented in
Ref.~\onlinecite{Neaton_et_al:2005} and all calculated structural
parameters agree well with experimental data. We note that the true
magnetic structure of \BFO\ shows some slight deviations from the
ideal G-type ordering, i.e. a long-wavelength spiral structure in the
bulk \cite{Sosnowska} and a slight canting of the magnetic moments in
thin films,\cite{Ederer/Spaldin:2005,Bai_et_al:2005} both caused by
spin-orbit coupling. However, these effects are rather small and not
relevant for the present study.

In the case of \BCO\ we also constrained the system to have rhombohedral $R3c$
symmetry and G-type antiferromagnetic ordering in our calculations, although
recently this system was reported to have monoclinic
symmetry.\cite{BiCrO3_Niitaka} This was done in order to be able to
systematically compare the properties of the series of compounds \BFO, \BFCO,
and \BCO.

\begin{table}
\begin{center}
\caption{Calculated lattice constant $a$, rhombohedral angle $\alpha$,
  volume $V$, and Wyckoff parameters for \BCO, \BFCO, and \BFO. The
  internal structural parameters refer to the Wyckoff positions 2$a$
  ($x$,$x$,$x$) for the cations and 6$b$ ($x$,$y$,$z$) for the anions
  (In case of the $R3$ structure the corresponding Wyckoff labels are
  1$a$ and 3$b$).
\label{table:compare}}
\begin{ruledtabular}
\begin{tabular}[c]{ l c c c c }
  & & \BCO & \BFCO & \BFO \\
\hline
$a$ [\AA] & & 5.43 & 5.47 & 5.50 \\
$\alpha$ [$^\circ$] & & 60.37 & 60.09 & 59.99 \\
$V$ [\AA$^{3}$] & & 114.47 & 116.86 & 117.86 \\
 & & & & \\
Bi & $x$ & 0.000 & 0.000/0.503 & 0.000 \\
Cr/Fe & $x$ & 0.228 & 0.226 (Cr)/0.732 (Fe) & 0.228 \\
O & $x$ & 0.546 & 0.547/0.045 & 0.542 \\
  & $y$ & 0.952 & 0.948/0.450 & 0.942 \\
  & $z$ & 0.407 & 0.405/0.898 & 0.368 \\
\end{tabular}
\end{ruledtabular}
\end{center}
\end{table}

The calculated lattice constants, rhombohedral angles, unit cell
volumes, and internal structural parameters for all three systems are
summarized in Table \ref{table:compare}. It can be seen that the
volume and lattice constants increase over the series \BCO, \BFCO,
\BFO, due to the larger radius of the high-spin Fe$^{3+}$ ion compared
with Cr$^{3+}$ (see also Sec.~\ref{ldau}). Likewise the rhombohedral
angle is reduced. The internal structural parameters, i.e. the
positions of the atoms within the unit cell are very similar in all
three systems. This partly reflects the fact that Bi is the ``active''
ion causing the ferroelectric distortion.

\subsection{\label{subsec:U}Effect of $U$ on the structure of \BFCO}
\label{ldau}

There is an interesting subtlety involved in the structural relaxation
of the \BFCO\ system, regarding the influence of the parameter $U$ on
the ground state structural parameters. Whereas LDA ($U$ = 0) leads
to a metallic low-spin state of the Fe$^{3+}$ cation (albeit not
changing the fact that the ground state is ferrimagnetic with $R3$
symmetry), the use of a moderate $U$
leads to an insulating solution containing high-spin Fe$^{3+}$. Due to
the substantial differences in the ionic radii of low-spin versus
high-spin Fe$^{3+}$ (the Shannon ionic radii are 0.55 and 0.65 \AA\
respectively \cite{Shannon}), this has a significant effect on the
lattice parameters. Using LDA for the structural relaxation we obtain
a lattice constant of $a$ = 5.30~\AA , a rhombohedral angle $\alpha$ =
61.23$^{\circ}$, and an equilibrium volume that is $\sim$ 7~\% smaller
than that obtained with LDA+$U$ and $U$ = 3~eV.

\begin{figure}
\begin{center}
\includegraphics[angle=0, width=0.5\textwidth]{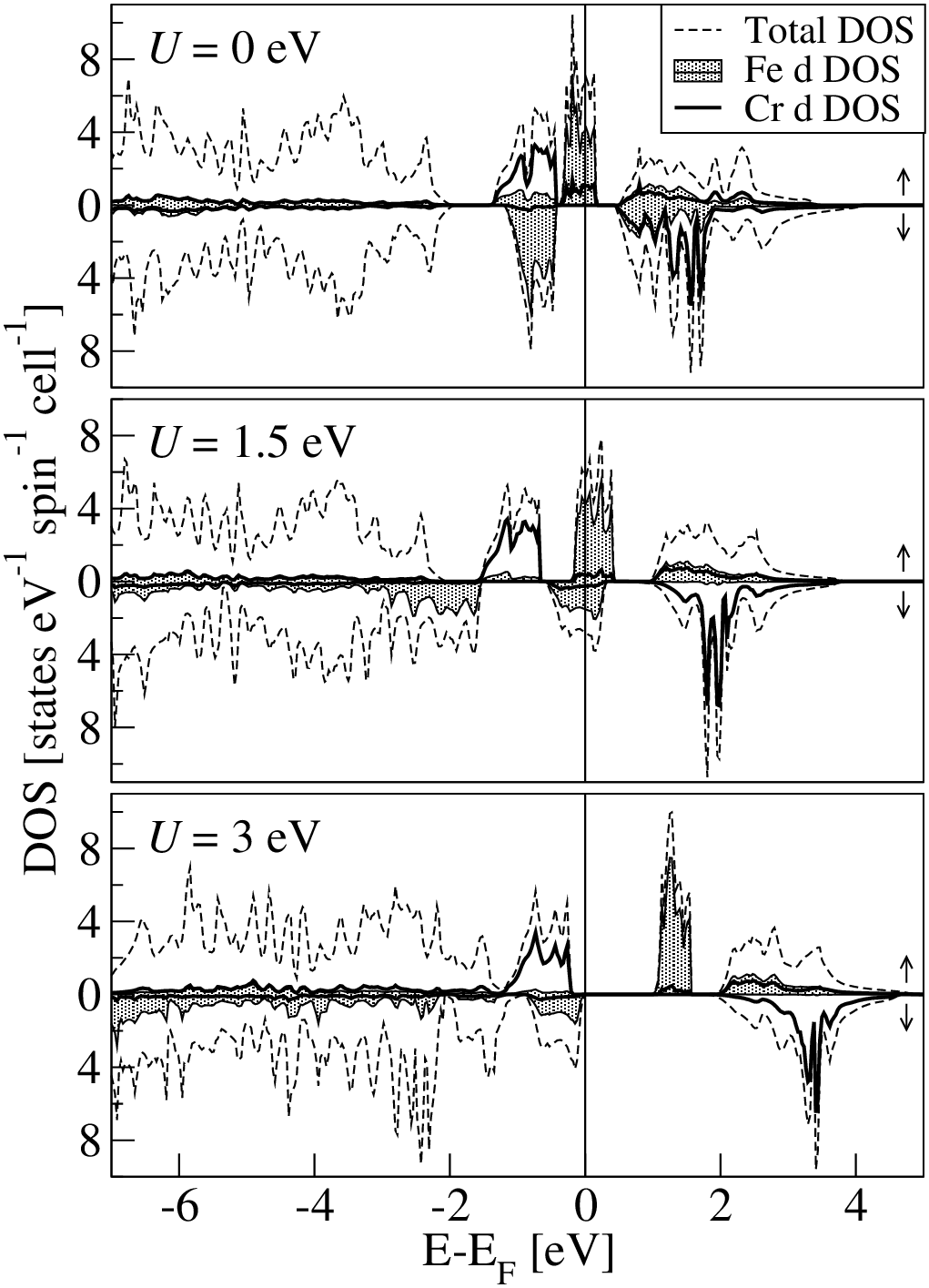}
\end{center}
\caption{Spin-resolved total and transition metal $d$ densities of
states (DOS) for $U$ = 0.0 eV (LDA), 1.5 eV and 3.0 eV. The total DOS
is indicated by the dashed lines, the Fe $d$ DOS by shading and the Cr
$d$ DOS by thick lines.
\label{fig:R3LDAU}}
\end{figure}

The evolution of the electronic structure with increasing $U$ is
illustrated in Figure~\ref{fig:R3LDAU}, which shows our calculated
spin-resolved densities of states for $U=$ 0 (the LDA result), 1.5~eV,
and 3 eV ($R3$ symmetry, FiM). To isolate differences in electronic
properties from structural effects, the same (LDA) structure is used
in all cases. The orbital-resolved densities of states are obtained by
projection onto a local angular-momentum basis and integrating up to a
radius of 1.4~\AA. In all cases, the spin of the Cr ion defines the
``up-spin'' direction, and its occupied $(t_{2g}^{\uparrow})^3$
manifold can be seen on the positive $y$-axis just below the Fermi
energy. In the LDA solution, the Fe $t_{2g}^{\downarrow}$ manifold is
completely filled, and the $t_{2g}^{\uparrow}$ states are partially
filled, leading to a finite density of states at the Fermi energy.
The calculated local magnetic moments (also calculated by integrating
up to a Wigner-Seitz radius of 1.4 \AA) on the Cr and Fe ions are
+2.5~$\mu_B$ and $-$0.6~$\mu_B$ respectively. These values are
slightly reduced from the ``ideal'' values of +3 and $-$1 expected for
Cr$^{3+}$ and low-spin Fe$^{3+}$ because of hybridization with the
oxygen band and the certain degree of arbitrariness in the definition
of the Wigner-Seitz spheres. Whereas the positions of the Cr states do
not change significantly as a function of $U$, the Fe states are
strongly affected by application of a small $U$ of 1.5 eV; the
$t_{2g}^{\downarrow}$ states shift down in energy and hybridize with
the oxygen $p$ band ($\sim$ 2 eV below the Fermi energy), and the
$e_{g}^{\downarrow}$ states shift down into the same energy region as
the $t_{2g}^{\uparrow}$ states, and become partially occupied. On
increasing $U$ to 3~eV, the Fe $e_{g}^{\downarrow}$ states become
fully occupied, the $t_{2g}^{\uparrow}$ states are completely empty,
and an insulating solution is obtained. The calculated local magnetic
moments are now +2.5 $\mu_B$ for the Cr$^{3+}$ ion (unchanged from the
LDA value) and $-$4.0 $\mu_B$ for the Fe$^{3+}$, again slightly
reduced from the ideal high-spin value of $-$5.0 $\mu_B$. Further
increase of $U$ does not change the electronic structure
qualitatively.

The above result shows that for \BFCO\ the application of the LDA+$U$
method induces a major change in the electronic structure which, in
turn, has a pronounced effect on the structural
parameters. Traditionally, the LDA+$U$ method has often been used in
combination with the atomic sphere approximation (ASA)
\cite{Andersen_ASA} which does not allow structural relaxation. Only
recently has the LDA+$U$ method also been applied to obtain structural
parameters, in general with considerable success (see
e.g. Refs.~\onlinecite{Neaton_et_al:2005},~\onlinecite{Rohrbach,GeOptLDAU2,GeOptLDAU1}).
Our results confirm that, in situations where the application of the
LDA+$U$ method can change the spin state of an ion, this latter
approach is essential for a meaningful result to be obtained.

Interestingly, if we relax the structural parameters using LDA, but fix the
total magnetic moment to be 8 $\mu_{B}$ (corresponding to high-spin states for
both Cr and Fe and a ferromagnetic configuration of all spins), the resulting
structure is very similar to that obtained by using LDA+$U$ with $U$ =
3~eV. This shows that the main effect of the on-site Coulomb repulsion in
\BFCO\ on the structure is the stabilization of the high-spin state of the
Fe$^{3+}$ cation. The larger volume of this high-spin state then determines
the structure. On the other hand, as long as the Fe$^{3+}$ cation is in its
high-spin state, the precise value of $U$ does not have a pronounced effect on
the structural parameters. As a consequence, the calculated LDA structure for
BiFeO$_3$ is very similar to the corresponding LDA+U structure, because in
both cases the transition metal ion adopts the same spin
configuration.\cite{Neaton_et_al:2005}

\subsection{\label{subsec:P}Calculated polarization}
\label{pol}

\begin{table}
\begin{center}
\caption{Spontaneous polarization $P_s$ calculated with the Berry
  phase method for all three systems, together with a simple estimate,
  obtained by summing up formal charges $Z_i$ times displacements
  $u_i$ for all ions $i$ in the unit cell (values in
  $\mu$C/cm$^2$). The ratio between these two quantities is also
  given.
\label{table:pol}}
\begin{ruledtabular}
\begin{tabular}{l c c c}
  & $P_s$ & $\sum_i Z_i u_i$ & $P_s/\sum_i Z_i u_i$ \\ 
\hline 
\BCO & 67 & 46 & 1.46 \\ 
\BFCO & 80 & 55 & 1.45 \\ 
\BFO & 95 & 64 & 1.48 \\
\end{tabular}
\end{ruledtabular}
\end{center}
\end{table}

Table \ref{table:pol} shows the spontaneous electric polarization
$P_s$ for the three systems, calculated with the Berry-phase method
(see Section~\ref{methods}), as well as an estimate for $P_s$,
calculated by multiplying the ferroelectric displacements, $u_i$, by
the formal ionic charges, $Z_i$, and summing up over all ions $i$. It
can be seen that, for both methods, the polarization increases from
\BCO\ to \BFCO\ to \BFO, reflecting a corresponding increase in
ferroelectric displacements of the ions. The polarization calculated
from the Berry phase is consistently larger than the simple estimate
by a factor of $\sim$ 1.5. This represents the well-known fact that
the effective charges in perovskite ferroelectrics are usually
significantly larger than the formal charges. (For example the
effective charges in ground state $R3c$ \BFO\ are 4.4, 3.5, and $-$2.5
for Bi, Fe, and O, respectively (see
Ref.~\onlinecite{Neaton_et_al:2005}), whereas the corresponding formal
charges are 3.0, 3.0 and $-$2.0). The nearly constant ratio between the
simple estimate and the Berry phase $P_s$, indicates that the
effective charges are comparable in all cases.

The polarizations of all three systems are relatively large and comparable to
that of PbTiO$_{3}$ (75 $\mu$C/cm$^{2}$).\cite{gavri} This suggests that
Bi-based perovskite ferroelectrics could be promising candidates for
alternative ``Pb-free'' ferroelectric and piezoelectric
devices.\cite{Baettig2005}

\subsection{Magnetic ordering temperatures and exchange couplings
  \label{section:M_T}}

To estimate the magnetic ordering temperature for the newly predicted
multiferroic \BFCO\, we determine the Heisenberg exchange constants
corresponding to nearest and next-nearest neighbor magnetic couplings
by comparing the total energies for different magnetic
configurations. From these coupling constants we then calculate the
Curie/N{\'e}el temperatures within the mean-field approximation (see
e.g. Ref.~\onlinecite{Morrish_book}). For comparison, we perform
analogous calculations for \BFO\ and \BCO. Note that the mean-field
approximation gives an upper limit for the actual ordering
temperature, since it neglects all fluctuations of the magnetic
moments from their expectation values.

To obtain the Heisenberg coupling constants for nearest and next
nearest neighbor coupling, we double the size of the unit cell (so
that it contains two formula units of \BFCO) and calculate the total
energies for several possible inequivalent collinear magnetic
configurations with different orientations of the magnetic moments on
different sites. We then map the calculated energies onto a simple
Heisenberg model, $E = - \frac{1}{2} \sum_{ij} J_{ij} S_i S_j$. Here
$S_i = \pm\frac{3}{2}$ on the Cr sites and $\pm\frac{5}{2}$ on the Fe
sites, and the sum is over all nearest and next-nearest neighbor
pairs. All calculated total energies can be reproduced accurately
using only nearest and next-nearest neighbor couplings.

\begin{figure}
\includegraphics[angle=0, width=0.5\textwidth]{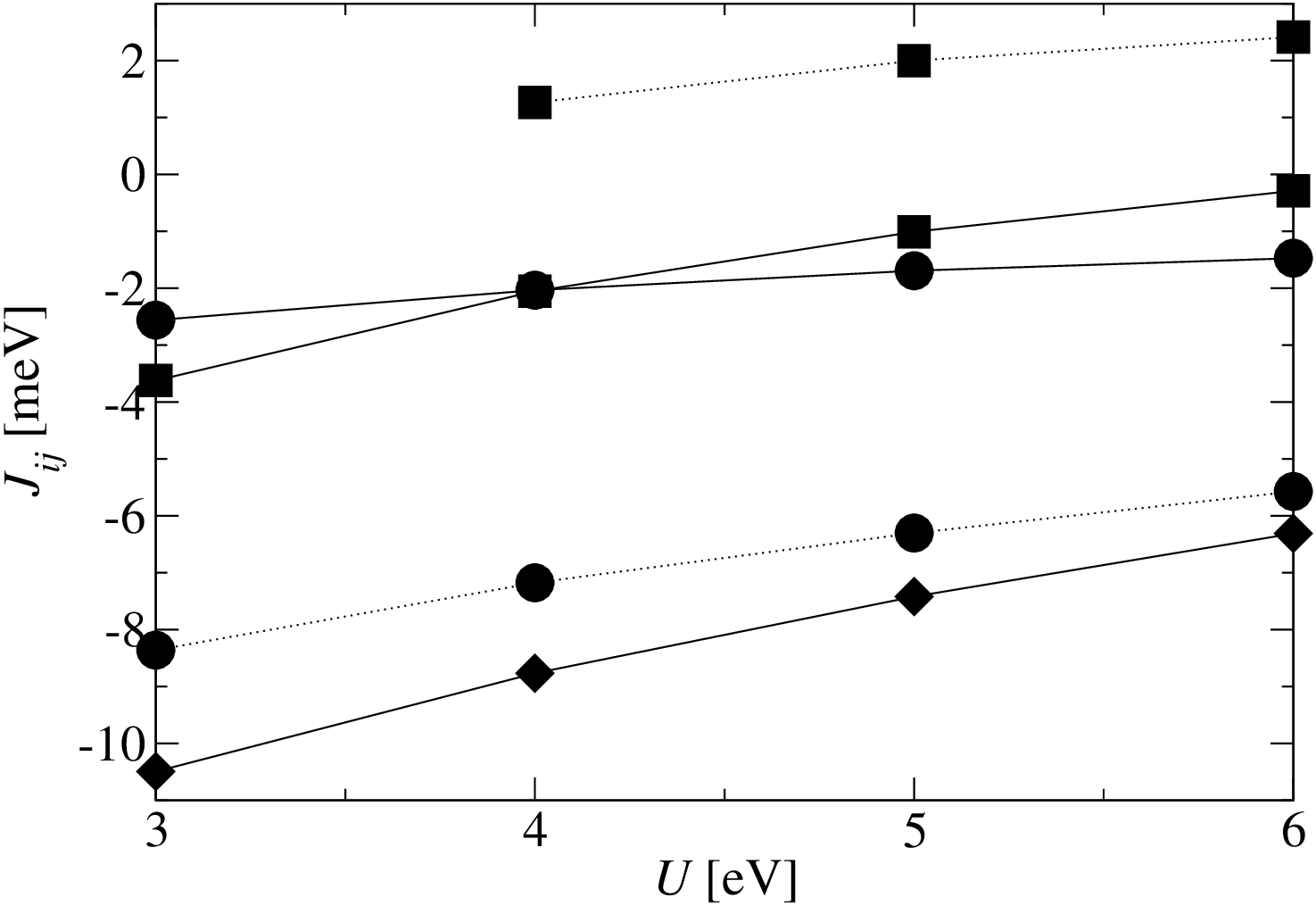}
\caption{Nearest-neighbor coupling constants, $J_{ij}$, for \BFO,
  \BFCO, and \BCO\ in their ground state structures as well as the
  corresponding coupling constants for \BFCO\ and \BCO\ in the ideal
  cubic perovskite structure as a function of $U$. Dashed lines are
  for the cubic structures, full lines are for the relaxed
  structures. Circles are for \BCO, squares for \BFCO, and diamonds
  for \BFO.}
\label{fig:couplings}
\end{figure}

In Refs.~\onlinecite{Solovyev:2002} and \onlinecite{Novak/Rusz:2005},
the coupling constants for Sr$_2$Fe$M$O$_6$ ($M$=W, Mo, Re) and for
BaFe$_{12}$O$_{19}$ were found to be rather sensitive to the value of
$U$. This reflects the dominant role of the superexchange interaction
in magnetic oxides, in which the antiferromagnetic coupling is
proportional to $b^2/U$, where $b$ is the transfer integral describing
hybridization effects.\cite{Anderson:1963} We therefore calculate all
coupling constants as a function of $U$ in the range 3~eV - 6~eV using
the structural parameters shown in Table~\ref{table:compare}. $J$ was
set to 1~eV for all systems. To separate out effects of changing
structure from differences in chemistry, we also calculate the
exchange couplings for ideal cubic perovskite \BCO, \BFO~ and \BFCO,
at the same average volume of
116.2~\AA$^{3}$. Figure~\ref{fig:couplings} shows the nearest neighbor
coupling constants for all three relaxed systems and for perfect cubic
\BCO\ and \BFCO\ as a function of $U$. (For cubic \BFO\ some magnetic
configurations are metallic up to $U$ = 6~eV. Since in the metallic
state different magnetic coupling mechanisms are present, we do not
include the values for cubic \BFO\ in our systematic comparison.)  For
both cubic and relaxed structures we obtain a reduction in the
strength of the antiferromagnetic superexchange as $U$ is increased
(Figure~\ref{fig:couplings}); indeed for cubic \BFCO\ the nearest
neighbor coupling even becomes ferromagnetic.

\begin{figure}
\includegraphics[angle=0, width=0.5\textwidth]{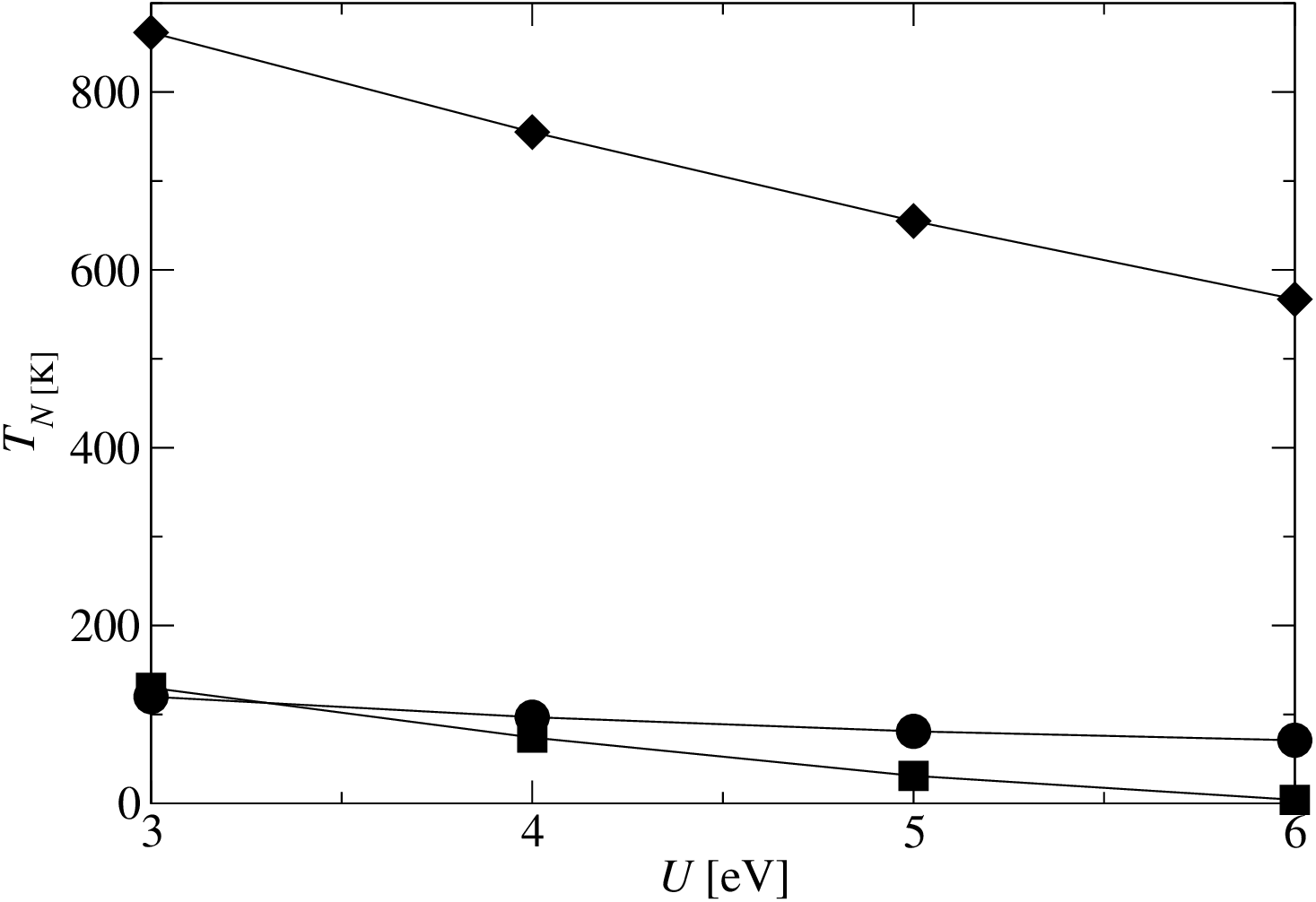}
\caption{Magnetic ordering temperatures for structurally relaxed \BCO,
  \BFCO, and \BFO\ as a function of the parameter $U$. Circles are for
  \BCO, squares for \BFCO, and diamonds for \BFO.}
\label{fig:temp}
\end{figure}

\begin{figure}[ht]
\includegraphics[angle=0, width=0.5\textwidth]{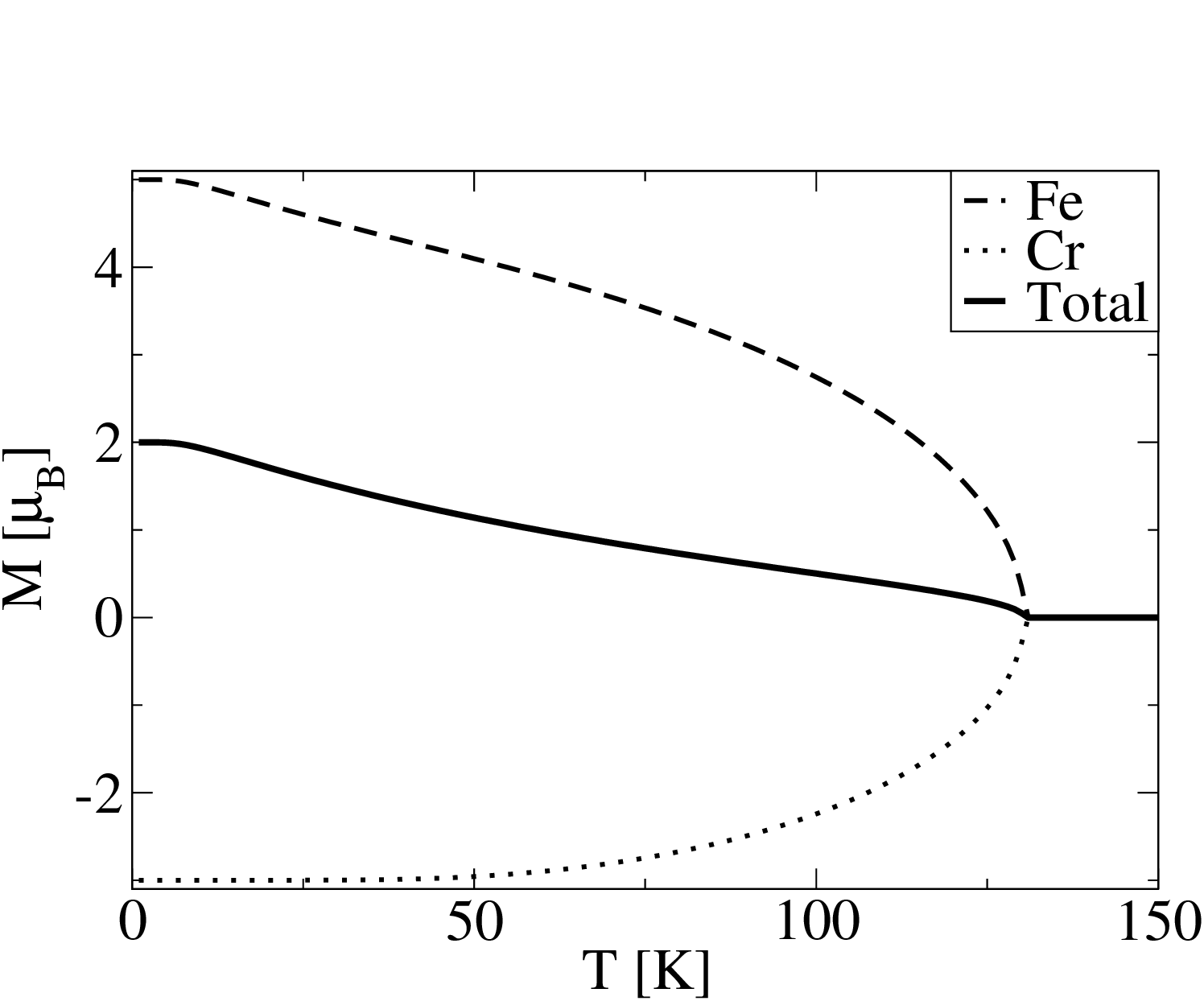}
\caption{Calculated temperature dependence of the magnetization for
  \BFCO\ ($U$ = 3~eV).  The dashed and dotted lines show the
  magnetizations of the Fe and Cr sublattices respectively.}
\label{fig:M_T}
\end{figure}

From the coupling constants $J_{ij}$ we calculate the magnetic
ordering temperature using the mean-field approximation (see
e.g. Ref.~\onlinecite{Morrish_book}). Figure~\ref{fig:temp} shows the
resulting magnetic ordering temperatures for all three
systems. Comparing the values of $T_N$ for \BFO\ with the experimental
value of 640~K shows that for $U \leq 5$~eV, as expected, the
mean-field approximation overestimates the magnetic ordering
temperature but gives the correct order of magnitude. For \BCO\ and
\BFCO\ the ordering temperatures are significantly smaller than for
\BFO, a fact that can be attributed to the presence of strong
$e_g$-$e_g$ coupling in the latter compound (see
Section~\ref{discussion}). Therefore, we expect that the magnetic
ordering temperature of the recently predicted multiferroic \BFCO\
will not exceed 100~K. The calculated $M(T)$ curve for \BFCO\ is shown
in Fig. \ref{fig:M_T}. We see that no ferrimagnetic compensation
temperature (where the two sublattice magnetizations cancel exactly)
is observed in the temperature range $0 < T < T_C$.

\section{Discussion}
\label{discussion}

According to Anderson \cite{Anderson:1963} the superexchange
interaction can be separated into two contributions: (i) \emph{kinetic
exchange} which is due to the mixing of the ligand-field orbitals used
to describe the spin quasiparticles; this term is proportional to
$b^2/U$ and is always antiferromagnetic, and (ii) \emph{potential
exchange} which represents the direct exchange interaction between
these ligand-field orbitals; this term is always ferromagnetic. Based
on this separation, Anderson gave the following simple guidelines for
estimating the signs and relative magnitudes of superexchange
interactions (note that the same outcomes are reached using the
related Goodenough-Kanamori rules\cite{Goodenough:Book}):
\begin{enumerate}
\item{Usually, the kinetic exchange is much stronger than the potential
exchange and dominates, leading to the predominantly antiferromagnetic
interactions found in magnetic oxides.}
\item{The kinetic exchange between $e_g$ electrons on different ions
  connected by a 180$^\circ$ metal-oxygen-metal bond is much stronger
  than the kinetic exchange between correponding $t_{2g}$ electrons,
  since the former is mediated by $dp\sigma$ bonds whereas the latter
  is mediated by weaker $dp\pi$ interactions.}
\item{In certain situations the kinetic exchange vanishes by symmetry
so that the remaining potential exchange leads to a small
ferromagnetic coupling. This occurs for example for a 90$^\circ$
superexchange coupling between $t_{2g}$ electrons and also in the case
of a 180$^\circ$ coupling between $e_g$ and $t_{2g}$ electrons.}
\item{Completely filled manifolds, with equal numbers of up- and down-spin
electrons give no net contribution to the superexchange interaction.}
\end{enumerate}
These rules explain, for example, the relative N\'{e}el temperatures
of the rock-salt structure transition metal monoxides shown in
Table~\ref{table:TCTN}. Ni$^{2+}$ has electron configuration
$(t_{2g}^{\uparrow})^3$, $(t_{2g}^{\downarrow})^3$,
$(e_{g}^{\uparrow})^2$. Therefore, only the two $e_g$ - $e_g$
interactions contribute to the net superexchange in the 180$^{\circ}$
Ni-O-Ni bond (the $t_{2g}^{\uparrow}$ - $e_g$ and
$t_{2g}^{\downarrow}$ - $e_g$ interactions are equal and opposite);
these are strongly antiferromagnetic, and the N\'{e}el temperature is
high. For Mn$^{2+}$, with electron configuration
$(t_{2g}^{\uparrow})^3$, $(e_{g}^{\uparrow})^2$, there is an
additional ferromagnetic $t_{2g}$ - $e_g$ direct exchange, which
weakens the antiferromagnetic coupling and lowers the N\'{e}el
temperature. CoO and FeO are intermediate.

\begin{table}
\begin{center}
\caption{N\'{e}el temperatures for different transition
metal oxides. All values except for LaFeO$_3$ were taken from
Ref.~\onlinecite{LB}).
\label{table:TCTN}}
\begin{ruledtabular}
\begin{tabular}{lc}
 Material & T$_{N}$ [K] \\
\hline
MnO & 118 \\
Fe$_{0.93}$O & 198 \\
CoO & 289 \\
NiO & 523 \\
\hline
LaCrO$_{3}$ & 282 \\
LaFeO$_{3}$ & 740 (Ref.~\onlinecite{Treves1965}) \\
\end{tabular}
\end{ruledtabular}
\end{center}
\end{table}

In \BCO\ the balance of the kinetic and potential exchange leads to a moderate
antiferromagnetic coupling between the $t_{2g}$ electrons on neighboring Cr
sites. This is stronger in the ideal cubic structure than in the distorted
$R3c$ structure because the perfect 180$^\circ$ Cr-O-Cr bond angles in the
high-symmetry case result in stronger kinetic exchange.

If half of the Cr$^{3+}$ ions in the cubic structure are replaced by
$d^5$ Fe$^{3+}$, the antiferromagnetic part of the nearest neighbor
coupling --- now between Fe$^{3+}$ ($d^5$) and Cr$^{3+}$ ($d^3$) ---
is drastically reduced and in fact the interaction becomes
ferromagnetic. This can be explained by the
additional {\it ferromagnetic} interactions between the $e_g$
electrons of the Fe$^{3+}$ and the $t_{2g}$ electrons of the Cr$^{3+}$
(see rule 3 above). When $U$ is increased, the remaining
antiferromagnetic kinetic exchange between the $t_{2g}$ electrons is
suppressed and the net interaction in the cubic structure becomes
ferromagnetic. Interestingly, in the relaxed structure of $R3$ \BFCO,
the nearest neighbor coupling between Fe$^{3+}$ and Cr$^{3+}$ is
weakly antiferromagnetic. This difference between the cubic and
relaxed structures can also be explained within the theory of
superexchange: when the structure is relaxed, the Fe-O-Cr bond angle
deviates from the perfect 180$^\circ$, allowing a certain degree of
mixing between Cr $t_{2g}$ and Fe $e_g$ states. This leads to a
nonzero kinetic exchange between these orbitals and a stronger
tendency for antiferromagnetic coupling. (Note that similar behavior
was discussed in Ref.~\onlinecite{Miura/Terakura} for the
La$_2$FeCrO$_6$ system.)

Finally, for \BFO\ the strong $e_g$-$e_g$ coupling due to the $dp\sigma$
bonding dominates and results in very strong antiferromagnetic nearest
neighbor coupling and a high magnetic ordering temperature.

A similar trend in magnetic ordering temperatures than discussed here
for the \BFO/\BFCO/\BCO\ system can also be observed for the closely
related LaFeO$_3$/La$_2$FeCrO$_6$/LaCrO$_6$ system (see Table
\ref{table:TCTN}). The ordering temperatures of the end members
LaFeO$_3$ and LaCrO$_3$ are comparable to the corresponding Bi systems
but appear to be slightly larger, which could be due to smaller
structural distortions in the non-ferroelectric La compounds. A Curie
temperature of 375~K and ferromagnetic order were reported for
(111)-layered La$_2$FeCrO$_6$,\cite{la2fecro6} but these results are
still under debate.\cite{Meijer:1998,Miura/Terakura} From our
discussion, ferromagnetic order could be expected for this system,
although the reported Curie temperature seems to be too high to be
explained by the weak ferromagnetic potential exchange caused by the
superexchange interaction.

From the above discussion it can be seen that high magnetic ordering
temperatures can be achieved in antiferro- or ferri-magnets by
exploiting the strong antiferromagnetic superexchange between $e_g$
electrons. The highest ordering temperatures in perovskite structures
with octahedral crystal field splittings can occur between two ions of
$d^8$ electron configuration, although this would of course lead to an
antiferromagnet with a net cancellation of magnetic
moments. Therefore, we propose that the best choice for a
ferroelectric with a net spontaneous magnetization is to combine $d^5$
and $d^8$ ions.  Here, provided that the crystal field splitting is
larger than the bandwidth, the material will be insulating, with a net
magnetic moment of 3 $\mu_B$ per formula unit.  While other
combinations, such as $d^8$ - $d^7$ would also give strong
superexchange coupling, the additional partially-filled sub-shells
reduce the probability of insulating behavior. In addition,
Jahn-Teller distortions and orbital ordering effects can lead to
antiferromagnetic coupling in one direction and ferromagnetic coupling
in another direction, which makes the simple rules outlined above more
difficult to apply.

Finally we discuss some possible combinations of elements that will
allow $d^5$ - $d^8$ coupling in the perovskite structure. First we
consider retaining Bi$^{3+}$ as the $A$-site cation in order to
exploit its well-understood stereochemically active lone pair to
induce the ferroelectric structural distortion. In this case an
average oxidation state of 3+ is required on the cation B site, which
limits the possible choices of magnetic ions considerably.
Co$^{4+}$/Ni$^{2+}$ would be a possible candidate for strong
$d^5$-$d^8$ coupling, but it is questionable if such a high oxidation
state can be achieved for Co.

Another possibility is to change the oxidation state of the anions,
for example by forming oxyfluorides. For example
Bi$_{2}$Mn(II)Ni(II)O$_{4}$F$_{2}$ would have the appropriate $d^5$
and $d^8$ electron configurations, and could give the added benefit of
an enhanced spontaneous electric polarization by strategic placement
of the F$^{-}$ anions. However, the magnetic coupling via F$^{-}$ is
weaker than via O$^{2-}$ and oxyfluorides are generally difficult to
prepare and explosive.

A better choice would be to change to a different $A$-site cation,
which is still lone-pair active, but which has a higher oxidation
state. For example a 4+ $A$-site cation would retain charge neutrality
with divalent d$^{5}$ Mn$^{2+}$ and d$^{8}$ Ni$^{2+}$ on the $B$
sites. In fact, several perovskite systems containing lone-pair active
Te$^{4+}$ on the $A$-site have been prepared in the
past,\cite{Kohn1976} with recent reserch focussing on
Se$_{1-x}$Te$_x$CuO$_3$, as a model system to study superexchange
interactions.\cite{Subramanian1999,Iniguez2005} However these systems
are usually strongly distorted due to the small size and the
stereochemically active lone pair of the Te$^{4+}$ ion, leading to
weak kinetic exchange and small magnetic ordering
temperatures.\cite{Kohn1976} Po$^{4+}$ is also a possibility, but
since it is radioactive it is less interesting for technological
applications.

Moving to the left in the periodic table, Pb$^{2+}$ and Tl$^+$ are
also lone pair active cations, and Pb$^{2+}$ of course is well known
as an $A$-site active cation in ferroelectric perovskites.  Here,
however, the average $B$-site oxidation state must be 4+, excluding
the possibility of a $d^8$ configuration. Therefore it is unlikely
that room-temperature Pb-based perovskite multiferroics will be
identified with this mechanism.  Even disregarding its toxicity, the
valence of Tl$^+$ is even more prohibitive.

In summary, although the spontaneous electric polarization of many
perovskite multiferroics persists to temperatures far above room
temperature, our analysis suggests that achieving large room
temperature magnetization will continue to be challenging. We suggest
Te$_2$MnNiO$_6$ as one possible candidate within the perovskite
structure, but it remains to be seen if the structural distortions in
this system are too large to result in strong superexchange
coupling. Our analysis indicates that other routes to magnetism, such
as the weak ferromagnetism believed to occur in
BiFeO$_3$,\cite{Ederer/Spaldin:2005} are worthy of further pursuit,
and that the search for multiferroics should be extended to structures
other than perovskite, in which stronger magnetic interactions might
be obtained.

\begin{acknowledgments}
Funding for this work was provided by the National Science
Foundation's {\it Chemical Bonding Centers} program, grant number
CHE-0434567, by the American Chemical Society's Petroleum Research
Fund, grant number 39440-AC5M, and by the MRSEC program of the
National Science Foundation, award number DMR00-80034. Discussions
with Rebecca Janisch and Claude Daul are gratefully acknowledged. NAS
acknowledges useful discussions with Dr. Pavel Novak during a visit to
the Czech Academy of Sciences supported by the International Center
for Materials Research, NSF grant number DMR-0409848.
\end{acknowledgments}


\end{document}